\documentclass[twocolumn,,amsmath,amssymb]{revtex4-1}

\usepackage{graphicx}
\usepackage{dcolumn}
\usepackage{bm}
\newcommand{\etal}{\emph{et al.}}
\newcommand{\be}{\begin{equation}}
\newcommand{\ee}{\end{equation}}
\newcommand{\bfig}{\begin{figure}}
\newcommand{\efig}{\end{figure}}

\usepackage{lineno}
\usepackage{lipsum}

\begin{document}      

\title{Phase tuning of multiple Andreev reflections of Dirac fermions and the Josephson supercurrent in Al-MoTe$_2$-Al junctions.
} 

\author{Zheyi Zhu$^{1}$}
\author{Stephan Kim$^1$}
\author{Shiming Lei$^2$}
\author{Leslie M. Schoop$^2$}
\author{R. J. Cava$^2$}
\author{N. P. Ong$^{1,\S}$}
\affiliation{
{$^1$Department of Physics, Princeton University, Princeton, NJ 08544, USA}\\
{$^2$Department of Chemistry, Princeton University, Princeton, NJ 08544, USA}
}

\date{\today}      
\begin{abstract}
{When a normal metal $N$ is sandwiched between two superconductors, the energy gaps in the latter act as walls that confine electrons in $N$ in a square-well potential. If the voltage $V$ across $N$ is finite, an electron injected into the well undergoes multiple Andreev reflections (MAR) until it gains enough energy to overcome the energy barrier. Because each reflection converts an electron to a hole (or vice versa), while creating (or destroying) a Cooper pair, the MAR process shuttles a stream of pairs across the junction. An interesting question is, given a finite $V$, what percentage of the shuttled pairs end up as a Josephson supercurrent? This fraction does not seem to have been measured. Here we show that, in high-transparency junctions based on the type II Dirac semimetal MoTe$_2$, the MAR leads to a stair-case profile in the current-voltage ($I$-$V$) response, corresponding to pairs shuttled incoherently by the $n^{th}$-order process. By varying the phase $\varphi$ across the junction, we demonstrate that a Josephson supercurrent ${\bf J}_{\rm s}\sim \sin\varphi$ co-exists with the MAR steps, even at large $V$. The observed linear increase in the amplitude of ${\bf J}_{\rm s}$ with $n$ (for small $n$) implies that ${\bf J}_{\rm s}$ originates from the population of pairs that are coherently shuttled. We infer that the MAR steps and the supercurrent are complementary aspects of the Andreev process. The experiment yields the percentage of shuttled pairs that form the supercurrent. At large $V$, the coherent fraction is initially linear in $n$. However, as $V\to 0$ ($n\gg 1$), almost all the pairs end up as the observed Josephson supercurrent.
}
\end{abstract}
 
\maketitle     


An active area of research is the investigation of proximity-induced pairing correlations in topological and other unconventional systems. Examples are graphene~\cite{Herrero,DuAndrei}, carbon nanotubes~\cite{Pillet}, point-contact or break junctions~\cite{Esteve,Pothier}, Josephson $\omega$-junctions~\cite{Strambini}, topological Bi nanowires~\cite{Bouchiat}, and systems exhibiting edge currents~\cite{Yacoby,Kim}. In these experiments, Andreev reflections and the associated subgap or bound states play central roles. Signatures specific to topological junctions arising from Andreev subgap states have been discussed by several groups~\cite{SanJose,vonOppen,Halperin}.

In the MAR process (Fig. \ref{figVmap}a), a right-moving electron in $N$ (red circle) is Andreev reflected at the right $N$-$S$ interface as a left-moving hole (white), which is, in turn, reflected as an electron at the left interface. If the voltage $V$ is finite, both excitations gain energy $eV$ with each traversal ($e>0$ is the elemental charge). Eventually, with $n$ traversals (the $n^{\rm th}$-order process), the excitation acquires enough energy to surmount the potential barrier. Two successive Andreev reflections shuttle one Cooper pair across the junction (green arrows). Previously, calculations have shown that subgap states in point-contact junctions give rise to a supercurrent when $V = 0$~\cite{Furusaki,Beenakker,Furusaki99}. We provide evidence that, even at large $V$, a supercurrent shuttled by the MAR process persists in large $S$-$N$-$S$ junctions.

Weak subgap, subharmonic features have long been observed in numerous experiments on (single) $S$-${\cal I}$-$S$ junctions (${\cal I}$ = insulator) and ascribed to various causes~\cite{Taylor,Marcus}. Their identification with MAR was made in Ref. \cite{Klapwijk}. Subsequently, microscopic calculations of MAR were compared with experiments on point-contact or break junctions~\cite{Averin, Bratus,Cuevas}.
Here we extend these pioneering experiments to a new regime, applying the powerful technique of phase tuning on high-transparency $SNS$ junctions using an asymmetric SQUID (superconducting quantum interference device) layout  ~\cite{Ouboter,Fulton,Barone,Esteve,Pothier,Bouchiat,vonOppen}. Our focus is on the MAR and the Josephson effect when $N$ is the Dirac-Weyl semimetal MoTe$_2$ (we stay above its critical temperature $T_{\rm c}$ = 100 mK~\cite{Kim}).

Flux-grown crystals of MoTe$_2$ exhibiting high residual resistivity ratios ($\sim$1 000) and mobilities (100 000 to 150 000 cm$^2$/Vs) were exfoliated in air into thin flakes of thickness~100 nm, and transfered onto a silicon substrate capped with a 90 nm-thick SiO$_2$ layer. We deposited Al wires on the surface of a flake to define four DC SQUIDs (details in Methods). As shown in Fig. \ref{figVmap}b, inset, the (sample) $S$-$N$-$S$ junction (1), with critical current $I_{\rm c}$, is fabricated in parallel with an auxiliary conventional $S$-${\cal I}$-$S$ tunnel junction (0) with critical current $I_{0}$ (where $S$ = Al, $N$ = MoTe$_2$ and ${\cal I}$ = Al$_2$O$_3$). The phases across the $S$-$N$-$S$ and $S$-${\cal I}$-$S$ junctions are $\delta$ and $\gamma$, respectively. In the 4 devices (1--4), the ratio $I_0/I_{\rm c}$ is $\ge 4$. The junction width $d$ equals 200, 300, 400 and 500 nm in devices S1, S2, S3 and S4, respectively.

When the applied current $I$ exceeds the SQUID's critical current $I_{\rm cS}$, the voltage $V$ rises steeply. We recorded both the $I$-$V$ curve and the differential resistance $dV/dI$ vs. $I$ in a magnetic field $B$ at temperatures $T$ from 0.135 to 1.1 K. The curves of $V(B, I)$ and $dV/dI(B, I)$ are reported as color maps in the $B$-$I$ plane (the two experimentally controlled quantities). 

In the regime $V=0$ ($I<I_{\rm cS}$), both phases $\delta$ and $\gamma$ are static and related by the constraint $\delta -\gamma = \varphi$, where $\varphi\equiv 2\pi BA/\phi_0$ is the flux-induced phase shift and $\phi_0 = h/2e$ is the superconducting flux quantum ($A$ is the loop area and $h$ is Planck's constant). For $V =0$, we maximize $I_{\rm cS}$ under the constraint on $\delta-\gamma$ and find that the auxiliary phase $\gamma$ is pinned close to $\pi/2$ if $I_0\gg I_{\rm c}$. The curve of $I_{\rm cS}$ vs. $\varphi$ then yields the CPR (current-phase-relation) curve of the SNS junction. From the CPR, we obtain $I_0 = 93 \,\mu {\rm A}$ and $I_{\rm c}= 29.6\,\mu$A (see Fig. \ref{figCPR} in Methods).

Our focus is on the finite-$V$ regime ($I>I_{\rm cS}$)~\cite{Ouboter}. 
At finite $V$, $\delta(t)$ and $\gamma(t)$ wind rapidly at the same rate ($\dot\delta = \dot\gamma$) with their difference fixed at $\varphi$. The finite voltage across the SQUID, given by $V = \hbar\langle \dot\gamma\rangle/2e$ ($\langle\cdots\rangle$ denotes time-averaged), drives a normal current $I_{\rm N}$ that flows parallel to the supercurrents $I_{s0}$ and $I_{s1}$ in the auxliary and sample junctions, respectively ($I = I_{s0}+I_{s1}+I_{\rm N}$).

To gain insight into the MAR, it is helpful to generalize the RSJ model. We assume that, at finite $V$, the supercurrent in the $S$-$N$-$S$ junction can be approximated by $I_{s1}\simeq I_{\rm A}(V)\sin\delta$, with a $V$-dependent amplitude $I_{\rm A}(V)$ (as $V\to 0$, $I_{\rm A}(0) = I_{\rm c}$). The total $t$-dependent current is then
\be
I(V,\varphi,t) = \frac{\hbar\dot\gamma(t)}{2e R_{\parallel}} + I_0\sin\gamma(t)+ I_{\rm A}(V)\sin(\gamma(t)+\varphi),
\ee
with $R_{\parallel}^{-1} = G = G_0+ G_1$ where $G_1$ is the shunt $S$-$N$-$S$ conductance arising from the MAR and $G_0$ is the remaining background conductance.

We consider the two extremal cases when $\varphi$ equals $\varphi_{+}$ (with $I_{s1}$ parallel to $I_{s2}$) and $\varphi_{-}$ (antiparallel). At $\varphi_\pm$, we have
\be
I(V,\varphi_{\pm},t) = \frac{\hbar\dot\gamma(t)}{2e R_{\parallel}} + [I_0\pm I_{\rm A}(V)]\sin\gamma(t),
\label{IV}
\ee
which reduces to the equation governing the phase dynamics in a single junction~\cite{Ouboter,Fulton,Barone}. 
(When the geometric inductance $L$ of the SQUID is negligible, $\varphi_+ = 0$ and $\varphi_- = \pi$. As discussed in Methods, a finite $L$ shifts $\varphi_{\pm}$ from these values~\cite{Barone}.)

We then have for the DC voltage
\be
V(\varphi_{\pm},I) = R_\parallel\sqrt{I^2 - I_{\rm cS}^2(\varphi_{\pm})}, \quad (I>I_{\rm cS}(\varphi_\pm))
\label{V0}
\ee
with $I_{\rm cS}(\varphi_{\pm}) = I_0 \pm I_{\rm A}(V)$. 
As shown below, allowing $I_{\rm A}$ and $R_{\parallel}$ to acquire a $V$ dependence yields a close description of the measured curves $V(\varphi_{\pm},I)$.

Figure \ref{figVmap}c shows $V(B,I)$ (measured in S1 at 135 mK) plotted in the $B$-$I$ plane. The black region ($V=0$) is bounded by the CPR curve $I_{\rm cS}(B)$. At fixed $B$, $V$ is observed to increase steeply once $I$ excceeds $I_{\rm cS}(B)$, approaching a linear increase at large $V$. 

In Fig. \ref{figVmap}d (main panel) we display a series of curves of $V$ vs. $I$ with $B$ as a parameter ($B\in [-24, \; -2]\;\mu$T). See Fig. \ref{figAppendix}a for curves with $B\in [0,\;42]\;\mu$T. A series of steps are clearly seen in $V(B,I)$.
As shown in the inset, they persist to $V= 100 \,\mu$V. The steps lead to narrow peaks in the derivative $dV/dI$ which follow a subharmonic sequence $V_n = V_{\rm max}/n$. In Fig. \ref{figVmap}e, we show that as $T$ is raised to 1.1 K, the peaks converge to zero.

A key feature emerges when we investigate the effect of phase tuning on the sequence of peaks. To see this, we plot the color map of $dV/dI(B,I)$ in the $B$-$I$ plane (Fig. \ref{figIBVB}a). Above the CPR curve $I_{\rm cS}(B)$, we see a sheaf of sharply defined sinusoidal curves that appear to peel off from the CPR curve. Since $B$ is linear in the phase $\varphi$, we infer that each peak is tracking a component of $I$ that varies sinusoidally with $\varphi$ -- a Josephson supercurrent.

We next show that each sinusoidal curve corresponds to an abrupt change in the MAR order ($n\to n-1$) at a fixed $V_n$. Using the $I$-$V$ curves in Fig. \ref{figVmap}d, we transform the vertical axis from $I$ to $V$. Under this transformation (Fig. \ref{figIBVB}a$\to$\ref{figIBVB}b), each sinusoidal curve in S1 collapses to a flat line. The voltage $V_n$ on each line fits the subharmonic sequence $V_n = V_{\rm max}/n$ for $n = 2,3,4 \cdots$ (Fig. \ref{figAppendix}c). The parameter $V_{\rm max}$ has the $T$-independent value 200 $\mu$V below 300 mK but decreases as $T\to T_c^{\rm Al}$ (1.2 K), consistent with $V_{\rm max} = 2\Delta/e$ where $\Delta$ is the energy gap of Al (Fig. \ref{figAppendix}b). These are key signatures of MAR~\cite{Klapwijk,Averin,Bratus,Cuevas}.

Increasing the junction spacing $d$ strongly damps both the critical current $I_{\rm c}$ and the amplitude of the sinusoidal curves. In Fig. \ref{figS1S4} we show color maps of $dV/dI(B,I)$ for the four devices S1$\cdots$S4 measured at 340 mK. Although, at 340 mK, $I_{\rm c}$ (defined as half the peak-to-trough excursion) decreases steeply with $d$ (see Fig. \ref{figAppendix}d in Methods), the supercurrent is observable well beyond $d$ = 500 nm at 135 mK. Hereafter, we focus on results from S1.

The striking staircase profile of $V(\varphi,I)$ (Fig. \ref{figVmap}d) provides a vital clue to the MAR charge transfer.
When $V$ satisfies $V_n\le V<V_{n-1}$, where $V_n = 2\Delta/ne$, the $n^{th}$-order process is dominant because of the divergent DOS at the gap edge in Al. In an ideal junction ($\tau =1$), the number of pairs shuttled is $n/2$ or $(n-1)/2$ for $n$ even or odd, respectively (a quasiparticle is also transferred for odd $n$). In both cases, the total charge shuttled is $ne$. Identifying each pair transfer as a conductance channel, we have $n$ channels when $V_n\le V<V_{n-1}$. At either end of the interval, the channel number abruptly changes by 1. 

Since $k_BT\ll \Delta/e$ at 135 mK, the normal current $I_{\rm N1}^{(n)}$ in the shunt conductance, $G_1^{(n)}$, derives overwhelmingly from pairs that are shuttled incoherently by MAR. As we show below (Fig. \ref{figFit}a), the shuttled pairs give rise to both a normal current $I_{\rm N1}^{(n)}= VG_1^{(n)}$ and a supercurrent; the former responds only to $V$ whereas the latter is sensitive to $\varphi$.

From the staircase profile we infer a simple expression for $G_1^{(n)}$. For $\tau = 1$, the abrupt change in conductance at $V=V_n$, 
$\Delta G_1 = G_1^{(n+1)} - G_1^{(n)}$, equals a constant $g$ (the conductance for one traversal). This implies $G_1^{(n)} = ng$.

In a non-ideal junction with $\tau<1$, classical scattering reduces the transmission probability by the amount ($1-\tau$) at each reflection. Instead of $G_1^{(n)}= ng$, we have
\be
G_1^{(n)} = g(1+\tau + \cdots + \tau^{n-1}) = g(1-\tau^n)/(1-\tau).
\label{G1}
\ee
As shown below, Eq. \ref{G1} leads to a quantitative description of $V(\varphi_\pm,I)$.

We turn next to the Josephson supercurrents $I_{s0}$ and $I_{s1}$.
The $n^{th}$-order sinusoidal curve in the $dV/dI$ map traces the variation of $I$ vs. $\varphi$ with $V=V_n$ (hereafter, we use $\varphi$ in place of $B$). The maximum and minimum values attained by the current are called $I_{n+}$ and $I_{n-}$, respectively. They occur at the extremal phases $\varphi_+$ and $\varphi_-$, respectively.

By Eqs. \ref{IV} and \ref{V0}, we have $I_{n\pm}^2 = (V_n/R_{\parallel})^2 + [I_0 \pm I_{\rm A}]^2$. Thus, the observed values $I_{n\pm}$ yield the observed amplitudes $I_{\rm A}^{(n)}$, viz.
\be
I_{\rm A,obs}^{(n)} = [I_{n+}^2 - I_{n-}^2]/(4I_0).
\label{IAobs}
\ee
These are plotted as black circles in Fig. \ref{figFit}a. Initially, $I_{\rm A,obs}^{(n)}$ increases linearly with $n$ but curves downwards when $n> 6$ (inset in Fig. \ref{figFit}a). To us, the initial $n$-linear growth of $I_{\rm A,obs}^{(n)}$ is persuasive evidence that the supercurrent derives from the population of Cooper pairs that are shuttled coherently.

At large $n$ ($V\to 0$), we expect $I_{\rm A,obs}^{(n)}$ to saturate since it cannot exceed the critical current $I_{\rm c}$ at $V=0$. We propose that saturation occurs because of the slight attenuation of the probability current at each reflection, analogous to Eq. \ref{G1}. Thus, $I_{\rm A}^{(n)}\sim 1 + \tau+\cdots + \tau^{n-1}$ which we write as
\be
I_{\rm A}^{(n)} = I_{\rm c}[1 -\tau^{n-1}].
\label{IA}
\ee

Using Eq. \ref{IA} to fit the data in Fig. \ref{figFit}a, we find that $I_{\rm c}$ = 27 $\mu$A, close to $I_{\rm c}$ = 29.6 $\mu$A obtained in the CPR. The agreement supports the reasoning behind Eq. \ref{IA}. The value of $\tau$ is found to be 0.924, slightly smaller than $\tau =$ 0.95 from the CPR fit (Eq. \ref{optimum}).

In Fig. \ref{figFit}a we also plot (as blue squares) the normal current in the sample junction $I_{\rm N1}^{(n)}$, the product of $V_n$ and $G_1^{(n)}$ (for $g$ we use the value from the fit in Eq. \ref{Vfit}).  
Despite the step-wise increase in $G_1$ with $n$, the decrease of $V\sim 1/n$ forces $I_{\rm N1}^{(n)}$ to decrease monotonically. By contrast, $I_{\rm A}^{(n)}$ (red and black symbols) increases monotonically before saturating at $I_{\rm c}$. Both currents reflect the series $1+\tau + \cdots +\tau^n$.

Heuristically, using Eqs. \ref{G1} and \ref{IA}, we can adopt the generalized RSJ model to describe the measured curves $V(\varphi_\pm, I)$ by the expressions
\be
V(\varphi_\pm, I) = \frac{\sqrt{I^2 - [I_0 \pm I_{\rm A}^{(n)}]^2}}{G_0+ G_1^{(n)}},
\label{Vfit}
\ee
where $n = {\rm Int}[V_{\rm max}/V]$. The fits are shown as red curves in Fig. \ref{figFit}b. A more sensitive test is to compare the fits to the total observed conductance $G_{\rm obs}(\varphi_\pm, I) \equiv [I^2 - (I_0 \pm I_{\rm A}^{(n)})^2]^{\frac12}V(\varphi_\pm, I)^{-1}$ (inset in Fig. \ref{figFit}b). As seen, both curves of $G_{\rm obs}(\varphi_\pm, I)$ fit well to Eq. \ref{Vfit}. From the fits, we find for $\varphi_-$ ($\varphi_+$)
\be
\tau = 0.93\; (0.92),\quad G_0 = 3.00 \;(2.94)\; {\rm S}, \quad g = 0.32 \; (0.20)\; {\rm S}.
\ee
The value of $\tau$ is in agreement with the fit to Eq. \ref{IA}. (For $\varphi_+$, a discontinuous jump of $V$ occurs at the threshhold current $I_{\rm cS}$. This implies a finite inertial term in the phase dynamics represented by a shunt capacitor $C$, which is neglected for simplicity.)

Using the phase-tuning technique, we have uncovered a direct relation between the voltage steps induced by MAR and the co-existing supercurrent. At large $V$ (small $n$), the well-resolved steps in $V$ correspond to a step decrease in the number of conductance channels for the pairs that are shuttled incoherently. Even at $n=2$, the coherently shuttled pairs in Device S1 produce a supercurrent that is detectable (Fig. \ref{figIBVB}a). In the opposite limit $V\to 0$, nearly all pairs are shuttled coherently. As a result, the amplitude $I_{\rm A}$ saturates to $I_{\rm c}$ obtained in the CPR (Fig. \ref{figFit}a). In between, the repeated reflections of the initial injected electron lays down the MAR tracks. However, both the normal current observed as steps in $G_1$ and the co-existing supercurrent arise from the Cooper pairs that are shuttled in its wake. 

In high-transparency junctions that display MAR processes up to high order, we expect a large fraction (possibly all) of the Josephson supercurrent to be comprised of pairs coherently shuttled by the MAR. The ability to measure how the amplitude $I_{\rm A}$ and other parameters ($n$, $\tau$, $g$, $\cdots$) vary with $V$ in phase-tuned junctions may lead to a more quantitative treatments of pairing correlations in the finite-$V$ regime, especially in unconventional platforms~\cite{SanJose,vonOppen,Halperin}.


\vspace{1cm}
\newpage

\section{Methods}

\subsection{Current definitions}
\noindent We provide a glossary of the current parameters and definitions.\\ \noindent
$I$ is the total current applied to the SQUID.\\ \noindent
$I_{\rm cS}$ is the critical current of the SQUID.\\ \noindent
$I_0$ is the critical current of the auxiliary junction.\\ \noindent
$I_{\rm c}$ is the critical current of the sample junction when $V=0$.\\ \noindent
$I_{\rm A}(V)$ is the $V$-dependent amplitude of the Josephson supercurrent at finite $V$.\\ \noindent
$I_1$ is the prefactor of the sample junction in Eq. \ref{CPR}.\\ \noindent
$I_{n\pm}$ are the maximum and minimum values of $I$ in the sinusoidal curve with $V=V_n$.\\ \noindent
$I_{\rm N1}$ and $I_{\rm N0}$ are the normal currents in the sample and auxiliary junctions, respectively, at finite $V$.\\ \noindent
$I_{\rm N}$ is the total normal current in the SQUID.\\ \noindent
$I_{s0}$ is the supercurrent in the auxiliary junction.\\ \noindent
$I_{s1}$ is the supercurrent in the sample junction.

\subsection{Device fabrication and measurement}
We used double-layer e-beam lithography, in combination with tilted-substrate thermal evaporation, to fabricate the $S$-${\cal I}$-$S$ junction. Initially, the substrate is spin-coated with MMA EL11 at 3 000 rpm for 30 s twice and baked at 175$^\circ$ C for 5 min, followed by spin-coating with PMMA 950 A07 at 4 000 rpm for 60 s and then baked at 175$^\circ$ C for 5 min. Next, the SQUID pattern was e-beam written using the Raith eLiNE writer with beam energy set at 30 kV, aperture at 10 $\mu$m and the dose level at $\sim$300 $\mu$C/cm$^2$. After developing in MIBK solution (MIBK: IPA=1:3) for 3 min and rinsing in IPA solution for 1 min, we fabricated a suspended bridge, using a half-dose beam to remove the underlying MMA layer, while keeping the suspended upper PMMA layer intact. The chip with the pattern defined is placed inside a thermal evaporator equipped with a tiltable stage, with vacuum at $\sim 5\times 10^{-7}$ mbar. To remove residual resist and several (oxidized) monolayers of MoTe$_2$, we exposed the chip to an RF Argon plasma \emph{in situ}. After cleansing, the first layer of Al (60 nm) is deposited at a rate of 10 $\AA$/s. Then a mixture of Ar/O$_2$ (10$\%$ O$_2$) is injected into the chamber for 30 min at $5\times 10^{-3}$ mbar to oxidize the Al. A second layer of Al (120 nm) is next deposited with the angle set at a new value to define an overlapping Al-AlO$_x$-Al junction under the suspended PMMA resist bridge. Finally, the device is immersed in acetone to wash off the extra Al layer. 

The $I$-$V$ measurements were performed in a top-loading wet dilution refrigerator (Oxford Instruments Kelvinox TLM400) with base temperature of 15--20 mK. The dc bias and ac excitation current is provided by Agilent 33220A function generator with a bias resistance of 100 k$\Omega$. After preamplification (NF LI-75A preamp.), the SQUID voltage $V$ was fed to a lock-in amplifier (Stanford Research SR830) for $dV/dI$ measurement, as well as a nanovoltmeter (Keithley 2182A) for $I-V$ measurement. Data above 300 mK were acquired in a Heliox Helium-3 cryostat with base temperature of 340 mK.

\subsection{The CPR curve}
We describe the fitting procedure to obtain the current-phase relation (CPR) curve from measurements on Device S1 at 135 mK. The CPR is the curve of the critical current $I_{\rm cS}(\varphi)$ bounding the dissipationless region where $V =0$. The total supercurrent is written as 
\be
I(B,\delta) =  I_0 \sin(\delta +\varphi) +  I_1\frac{\sin\delta}{\sqrt{1-\tau\sin^2(\delta/2)} },
\label{CPR}
\ee
with $\varphi = 2\pi \phi/\phi_0$ where $\phi$ is the total flux piercing the SQUID and $\phi_0=h/2e$ is the superconducting flux quantum. For the $SNS$ junction, we adopted the CPR expression ${\cal F}(\delta)\equiv \sin\delta/\sqrt{1-\tau\sin^2(\delta/2)}$ appropriate for subgap states in junctions with high transparency ($\tau\to 1$). Its prefactor $I_1$ is proportional to $I_{\rm c}$. The total flux $\phi$ is the sum of the applied flux $\phi_{\rm a}$ ($\phi_{\rm a}=BA$) and the Amperean flux ``$LI$'' produced by supercurrents flowing around the SQUID loop. We have, when $V=0$,
\be
\phi = \phi_a - L_0I_0\sin(\delta + \varphi) + L_1 I_1{\cal F(\delta)}.
\label{phitot}
\ee
The partial equivalent inductances $L_0$ and $L_1$ represent the partitioning of $L$ into contributions from the two branches ($L = L_1 + L_2$)~\cite{Fulton}. 

In our device layout, we have $L_0\simeq L_1 \simeq L/2$. We estimate the total inductance $L$ as that of a square loop (of sides 6 $\mu$m and linewidth 1 $\mu$m) to be 8.2 pH. (This is close to the value estimated from the shift of the mimima of each sinusoidal curve in the color map of $dV/dI$ (Fig. \ref{figIBVB}a) $dI/dB$ which yields $L\sim$ 9 pH as shown below.)

With the total flux given by Eq. \ref{phitot}, the optimization of the fit by analytical means is highly unstable. Instead, we used the following numerical procedure. Starting with seed values for the 3 unknowns, $\tau$, $I_0$ and $I_1$, we allow $\varphi$ and $\delta$ to range over the full parameter space $\varphi \in [0, 2\pi]$, $\delta \in[0,2\pi]$. The magnitude of $I(\varphi, \delta)$ at each point $(\varphi, \delta)$ (computed using Eqs. \ref{CPR} and \ref{phitot} with $L$ = 8.2 pH) defines a surface $I(\varphi,\delta)$. When we project the surface onto the $\varphi$-$I$ plane, its upper boundary (maximum $I$) gives the CPR curve (using Eq. \ref{phitot} to convert $\varphi$ to $B$). The deviation of this calculated CPR from the observed CPR yields an error function (a scalar map in the space of $\tau, I_0, I_1$). By iteration, we converge rapidly to the optimal fit values. 
\be
 \tau= 0.95, \quad I_0= 93 \,\mu {\rm A}, \quad I_1 = 24 \,\mu {\rm A}.
\label{optimum}
\ee

In the $V=0$ regime, we define the critical current $I_{\rm c}$ of the $S$-$N$-$S$ junction as one-half the trough-to-peak excursion of the CPR curve. From the red circles in Fig. \ref{figCPR}, we get $I_{\rm c}\simeq$ 29.6 $\mu$A. Hence the prefactor $I_1 \simeq 0.81\; I_{\rm c}$.
 
 \subsection{Data in Device S1}
Figure \ref{figAppendix}a shows curves of $V(B,I)$ in Device S1 for $0< B< 42\;\mu$T (complements the curves in Fig. \ref{figVmap}a). Figure \ref{figAppendix}b shows that $V_{\rm gap}$ in S1 has a $T$ dependence consistent with the gap order parameter in Al. In \ref{figAppendix}c, the plot verifies that $V_n$ measured at the flat lines in Fig. \ref{figIBVB}b satisfies $V_n = V_{\rm max}/n$. Both Panels (b) and (c) support the identification of the steps in $V(B,I)$ with MAR. In Fig. \ref{figAppendix}d, we show the decrease in $I_{\rm c}$ with junction spacing $d$ ($I_{\rm c}$ is inferred from the CPR curves). At $T$= 135 mK, the decrease is gradual, but at 340 mK, $I_{\rm c}$ decays steeply for $d>$200 nm.

\subsection{RSJ model of single junction}
First we apply the RSJ model to the $SNS$ junction 1 in isolation. Treating it as an overdamped Josephson junction (JJ) (setting the inertial term $C=0$), we have a resistance $R$ in parallel with the JJ. When the applied current $I>I_{\rm c}$, winding of the phase $\theta$ leads to a time-dependent voltage $V(t) = \hbar \dot{\theta}(t)/2e$. Adding the Josephson supercurrent to $V(t)/R_1$, the total current $I$ is 
\be
I = \hbar \dot{\theta}/2e R + I_{c}\sin\theta.
\ee
The differential equation may be integrated to solve for the winding rate
\be
\dot{\theta}(t) = \frac{p^2-1}{p}\left(\frac{\cos\theta}{\cos\frac{\tau}{2}\sqrt{p^2-1}}\right)^2
\label{winding}
\ee
where $\tau = (2I_ceR/\hbar)t$, and $p = I/I_c$. The winding rate $\dot{\theta}(t)$ is a periodic function of $t$ with narrow peaks separated by the period $T$ given by
\be
T = \frac{2\pi}{\sqrt{I^2 - I_{\rm c}^2}}\frac{\hbar}{2eR}.
\label{T}
\ee

Time-averaging the winding rate $\langle \dot{\theta}\rangle$, we obtain for the DC voltage 
\be
\langle V\rangle = \frac{\hbar}{2e}\oint \dot{\theta} \frac{dt}{T} =  R \sqrt{I^2 - I_{\rm c}^2}
\ee

\subsection{Asymmetric SQUID with finite inductance $L$}
In the finite-$V$ regime, we have for the total current
\be
I =  \frac{\hbar\dot{\gamma}}{2e R_\parallel} + I_0\sin\gamma + I_{\rm A}(V)\sin\delta,
\label{Itot2}
\ee
where we assume that ${\cal F}(\delta)\simeq \sin\delta$ and replace $I_{\rm c}$ with the amplitude $I_{\rm A}(V)$. The color plot in Fig. \ref{figIBVB}a shows that the sinusoidal assumption is reasonable except very close to the CPR curve ($V\to 0$). In applied $B$, the phases $\gamma$ and $\delta$ remain related by the total flux $\phi$:
\be
\gamma - \delta = \varphi = \frac{2\pi}{\phi_0}\phi.
\label{phishift}
\ee

In the presence of the normal currents $I_{\rm N0}=V/R_0$ and $I_{\rm N1}=V/R_1$ driven by $V$, Eq. \ref{phitot} changes to
\be
\phi = \phi_a - L_0I_0(\sin\gamma + \frac{I_{\rm N0}}{I_0}) + L_1 I_{\rm A}(\sin\delta + \frac{I_{\rm N1}}{I_{\rm A}}).
\label{phinew}
\ee

Defining $\varphi_a = (2\pi/\phi_0)\phi_a$ and eliminating $\phi$ using Eq. \ref{phishift}, Eq. \ref{phitot} becomes
\be
\varphi_a = \gamma - \delta + \beta_0 (\sin\gamma +\frac{I_{\rm N0}}{I_0}) - \beta_1 (\sin\delta + \frac{I_{\rm N1}}{I_{\rm A}}).
\label{varphia}
\ee
with the beta parameters $\beta_0 = 2\pi I_0 L_0/\phi_0$ and $\beta_1 = 2\pi I_{\rm A} L_1/\phi_0$. In Device 1, $\beta_0\simeq 1.17$ and $\beta_1\simeq 0.38$.

Close to the threshold ($V\to 0$), the nonlinear equations Eqs. \ref{Itot2} and \ref{varphia} require a numerical solution, but the solutions are difficult to relate to measurable quantities. We restrict attention to the voltage regime $V> V_{\rm max}/n_{\rm max}$ where $n_{\rm max}= 11$  is the highest MAR order resolved. In this limit, $I_{\rm N1}\simeq 0.9 I$ and $I_{\rm N0}\simeq 0.1 I$, i.e. $90\%$ of the applied current flows through the SNS junction as a normal current. Then Eq. \ref{varphia} simplifies to
\be
\varphi_a \to \gamma - \delta  + \beta_0\sin\gamma -\beta_1\frac{I_{\rm N1}}{I_{\rm A}}.
\ee
Substituting this into Eq. \ref{Itot2}, we see that at $\varphi_+$, when the supercurrents are parallel ($I_{s0}\parallel I_{s1}$ ) the corresponding $\varphi_a$ is
\be
\varphi_{a+} = \beta_0\sin\gamma - \beta_1 \frac{I_{\rm N1}}{I_{\rm A}}.
\label{shift}
\ee
This shows that the maximum $I_+^{(n)}$ in a sinusoidal curve in Fig. \ref{figIBVB}a is shifted from 0 by an amount linear in $I_{\rm N1}=0.9 I$. Similarly, $\varphi_-$ is shifted from $\pi$ by the same amount as in Eq. \ref{shift}. 

In the large $I$ limit, Eq. \ref{shift} further simplifies to
\be
\varphi_{a+} \simeq \frac{2\pi }{\phi_0}\frac{L V}{2R_1}.
\label{device}
\ee
The $V$-linear shifts of $\varphi_\pm$ are clearly observed in the color map of $dV/dI$ (Fig. \ref{figIBVB}a).

The primary effect of $L$ is to induce a phase shift in the sinusoidal curves without affecting the amplitude. In our fits to Eq. \ref{Vfit}, $L$ does not enter explicitly because we use the experimentally observed shifts to locate $\varphi_{a+}$ and $\varphi_{a-}$, respectively.

\vspace{1cm}
\centerline{* ~~~ * ~~~  *}

\newpage
\vspace{1cm}\noindent
$^{\S}$Corresponding author email: npo@princeton.edu\\

\vspace{5mm}\noindent
{\bf Acknowledgements} \\
S. K. and the experiments performed at milliKelvin temperatures were supported by an award from the U.S. Department of Energy (DE-SC0017863). The crystal growth, led by R. J. C., L. M. S. and S. L., was supported by MRSEC award from the U.S. National Science Foundation (NSF DMR-2011750). Z. Y. Z. and N. P. O. were supported by the Gordon and Betty Moore Foundation's EPiQS initiative through grant GBMF4539.

\vspace{3mm}
\noindent
{\bf Author contributions}\\
The experiment was designed by Z.Y.Z. and N.P.O. and carried out by Z.Y.Z. and S.K. on successive generations of MoTe$_2$ crystals grown by S.L., L.M.S. and R.J.C. Analysis of the data was done by Z.Y.Z. and N.P.O. The manuscript was written by N.P.O. and Z.Y.Z.

\vspace{3mm}
\noindent
{\bf Additional Information}\\
Supplementary information is available in the online version of the paper.
Correspondence and requests for materials should be addressed to N.P.O.

\vspace{3mm}
\noindent
{\bf Competing financial interests}\\
The authors declare no competing financial interests.

\newpage
\noindent
{\bf Figure Captions}\\
\noindent
{\bf Figure 1}\\\noindent
Multiple Andreev reflections in the asymmetric SQUID layout with the sample $S$-$N$-$S$ junction based on MoTe$_3$. Panel (a) shows a sketch of the $5^{th}$-order MAR process. Right-moving electrons (red circles) are Andreev reflected as left-moving holes (white circles). After 5 traversals, the excitation gains sufficient energy to scale the gap barrier. In the process 2 pairs (green arrows) are shuttled across. The density of states (DOS) in Al are shaded grey. Panel (b) shows four devices fabricated on an exfoliated crystal of MoTe$_2$ (gray area). The sketch in the upper inset shows the $S$-$N$-$S$ ($S$-${\cal I}$-$S$) junction on the right (top) branch. The applied flux $\phi_{\rm a}=BA$ pierces the enclosed rectangle (see scale bar). In Panel (c) the color map is comprised of 150 $I$-$V$ curves with spacing $\Delta B = 2\, \mu$B (color scale in vertical bar) measured in S1 at $T$ = 135 mK. At fixed $B$, $V$ increases steeply once $I$ exceeds the SQUID critical current $I_{\rm cS}(B)$ (the CPR curve). If $V$ is held fixed, the current varies periodically with vs. $B$ up to $V\sim$ 80 $\mu$V (thin black curves are contours of $V(B,I)$). Panel (d) (main panel) shows the $I$-$V$ curves with $B$ varying from -24 $\mu$T to -2 $\mu$T in steps of 2 $\mu$T. Subharmonic steps ($n$ as indicated) are visible up to $V$ = 105 $\mu$V (see inset). Peaks in $dV/dI$ at the steps occur at $V_n = V_{\rm max}/n$ ($n=2,3,\cdots$). Panel (d) shows traces of $dV/dI$ vs. $V$ for $0.34 < T< 1.10$ K. As $T\to T_{\rm c}^{\rm Al}$, $V_{\rm max}\to 0$ in accord with MAR.

\vspace{2mm}\noindent
{\bf Figure 2}\\\noindent
Color maps of the differential resistance $dV/dI(B,I)$ measured in S1 at 135 mK, displayed in the $(B,I)$ plane (Panel a) and the $(B,V)$ plane (b) with scale bar on right. In Panel a, the black area ($dV/dI = 0$) is bounded by the CPR curve $I_{\rm cS}(B)$. Above the CPR, the series of sinusoidal curves trace the variation of narrow peaks in $dV/dI$ vs. $B$. In Panel b, the vertical axis is transformed to $V$ using the $I$-$V$ curve at each $B$. Each sinusoidal curve collapses to a flat line with $V$ fixed at the MAR $V_n = V_{\rm max}/n$.

\vspace{2mm}\noindent
{\bf Figure 3}\\\noindent
Comparison of color maps of $dV/dI(I,B)$ measured in S1--S4 at $T$ = 340 mK. The junction spacing $d$ is 200, 300, 400, 500 nm in S1 to S4, respectively. As $d$ increases, the supercurrent amplitude $I_{\rm c}$ inferred from the CPR decreases rapidly at 340 mK (see Fig. \ref{figAppendix}d in Methods). Unlike at 135 mK, $dV/dI$ is not strictly zero below the CPR curve because of thermally excited quasiparticles.

\vspace{2mm}\noindent
{\bf Figure 4}\\\noindent
Fits to the supercurrent amplitude $I_{\rm A,obs}^{(n)}$, the $I$-$V$ curves and total conductance $G_{\rm obs}$ in Device S1.
Panel (a): The observed supercurrent amplitude $I_{\rm A,obs}^{(n)}$ vs. $n$ (Eq. \ref{IAobs}, black circles) compared with $I_{\rm A}^{(n)}$ calculated using Eq. \ref{IA} (red circles). The best fit is obtained with $I_{\rm c}$ = 27 $\mu$A and $\tau = 0.924$. The normal current $I_{\rm N1}^{(n)}= V_n G_1^{(n)}$ is plotted as blue squares.
The data for $I_{\rm A,obs}^{(n)}$ and fit are shown in expanded view in the inset. 
In Panel (b) (main panel), we compare the measured $V(\varphi_\pm, I)$ (black curves) with the best fit to Eq. \ref{Vfit} (red curves). The integer $n$ at each step is indicated. For clarity, the curve $V(\varphi_+,I)$ is shifted upwards by 12.5 $\mu$A. The inset shows the same fits compared against the observed conductances $G_{\rm obs}(\varphi_\pm,I)$ defined below Eq. \ref{Vfit}, which provide a more sensitive test for the fits.

\vspace{2mm}\noindent
{\bf Figure S1}\\\noindent
Numerical procedure for fitting the CPR equation Eq. \ref{CPR} to the curve of $I_{\rm cS}(B)$ measured in S1 at 135 mK. The $x$ axis is first converted to the corrected field $B_{\rm cor}$ using Eq. \ref{phitot}. The set of values for $I(B_{\rm cor},\delta)$ is calculated from Eq. \ref{CPR} allowing $\delta$ and $\varphi$ to vary over $[0,2\pi]\times[0,2\pi]$. These values are projected onto the $I$-$B_{\rm cor}$ plane. The difference between the maximum value taken by $I$ and the measured CPR (red circles) defines the error function. We iteratively reduce the error function to force convergence to the optimal values Eq. \ref{optimum}. The wavy motif in gray is an artefact of the procedure with no physical significance.

\begin{figure*}[t]
\includegraphics[width=14 cm]{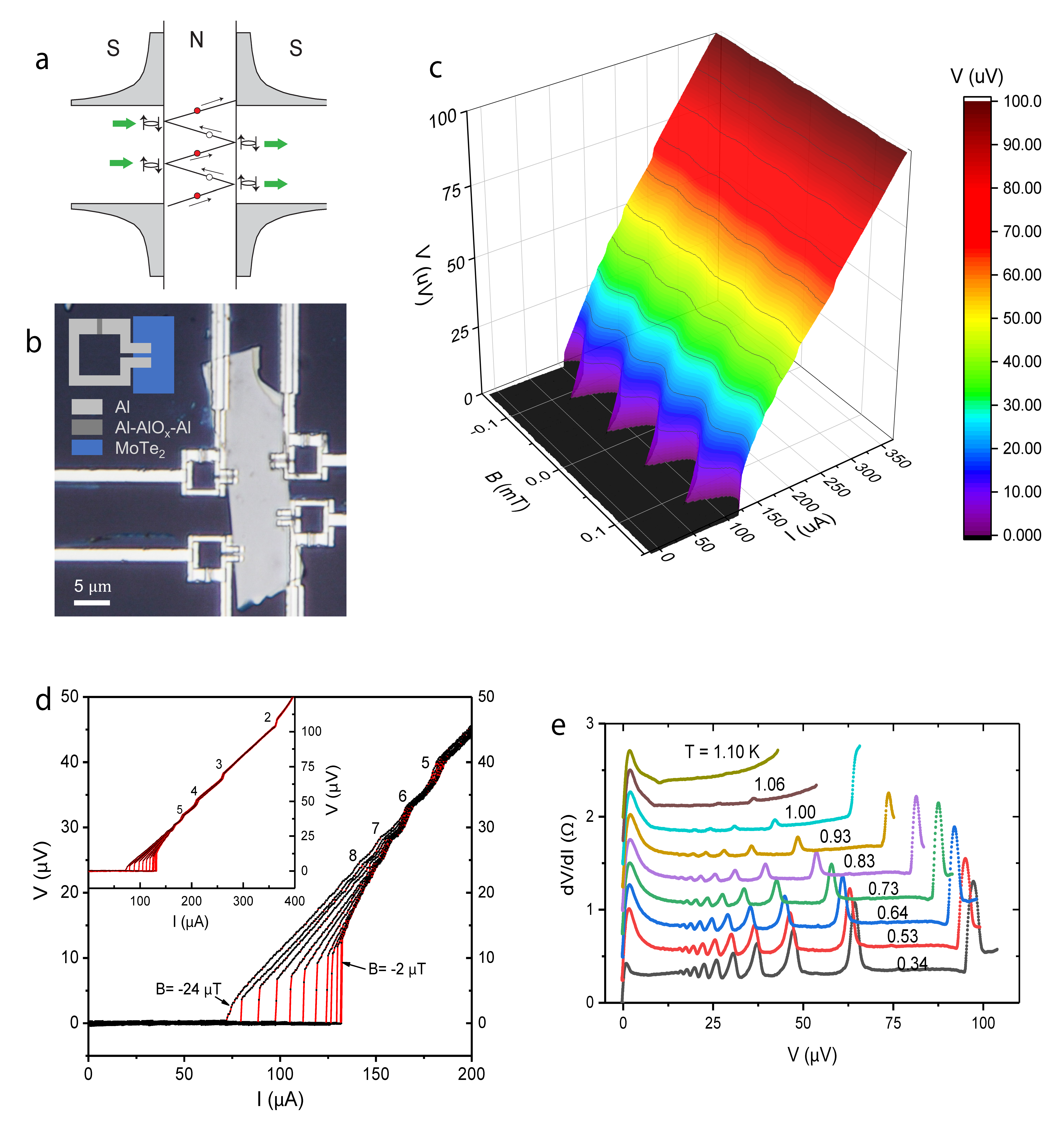}
\caption{\label{figVmap} 
}
\end{figure*}

\begin{figure*}[t]
\includegraphics[width=16 cm]{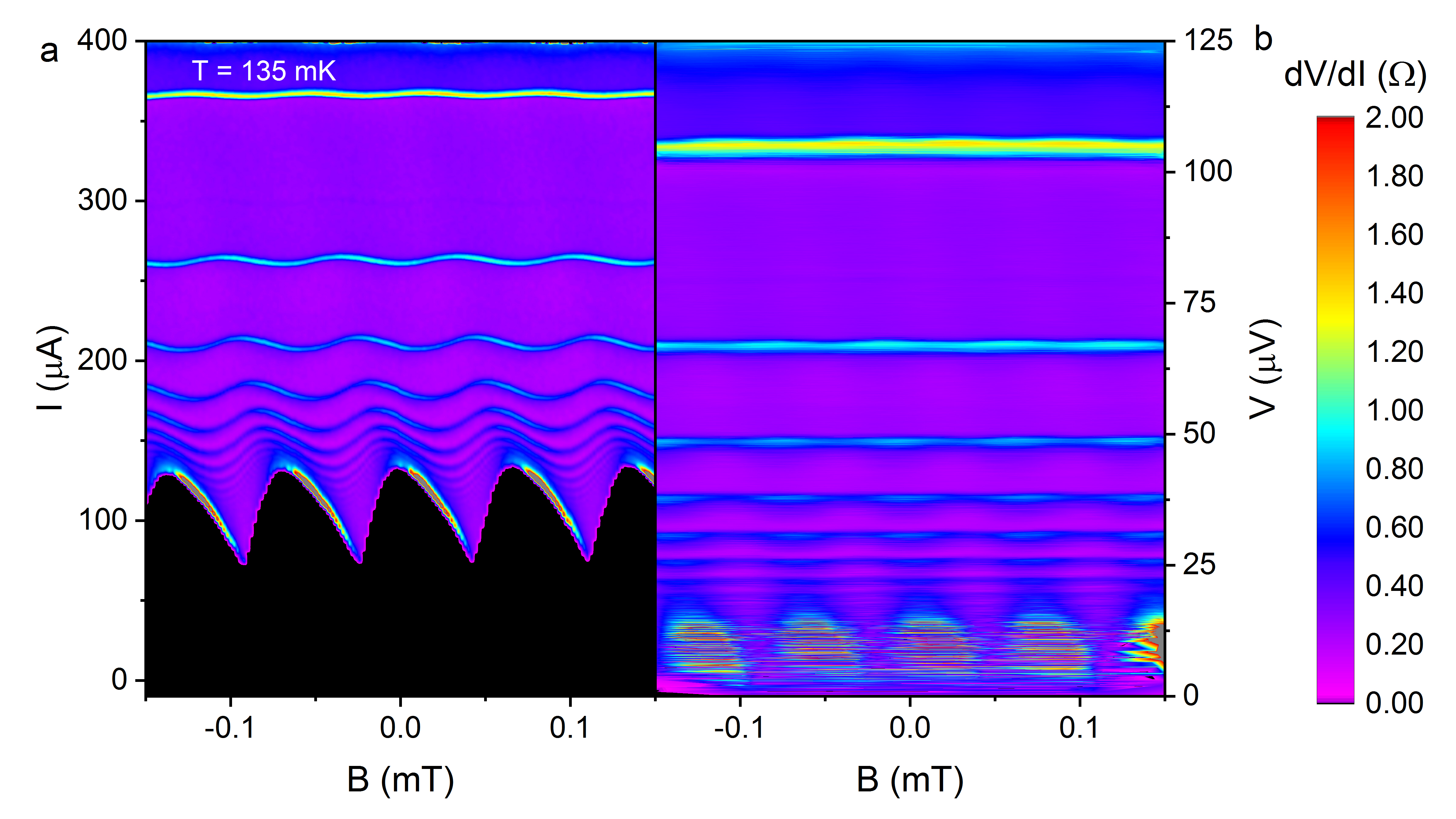}
\caption{\label{figIBVB} 
}
\end{figure*}

\begin{figure*}[t]
\includegraphics[width=16 cm]{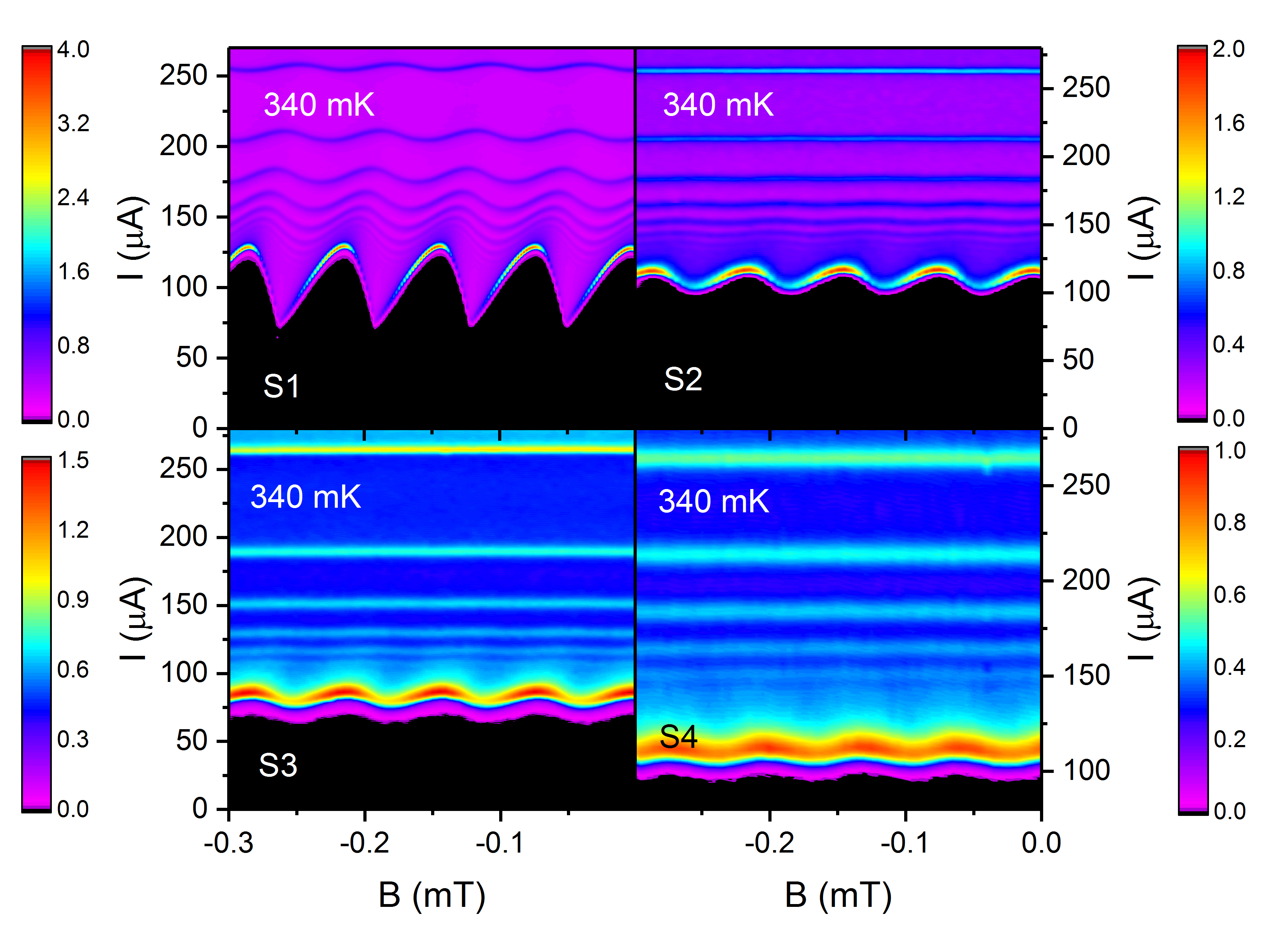}
\caption{\label{figS1S4} 
}
\end{figure*}

\begin{figure*}[t]
\includegraphics[width=18 cm]{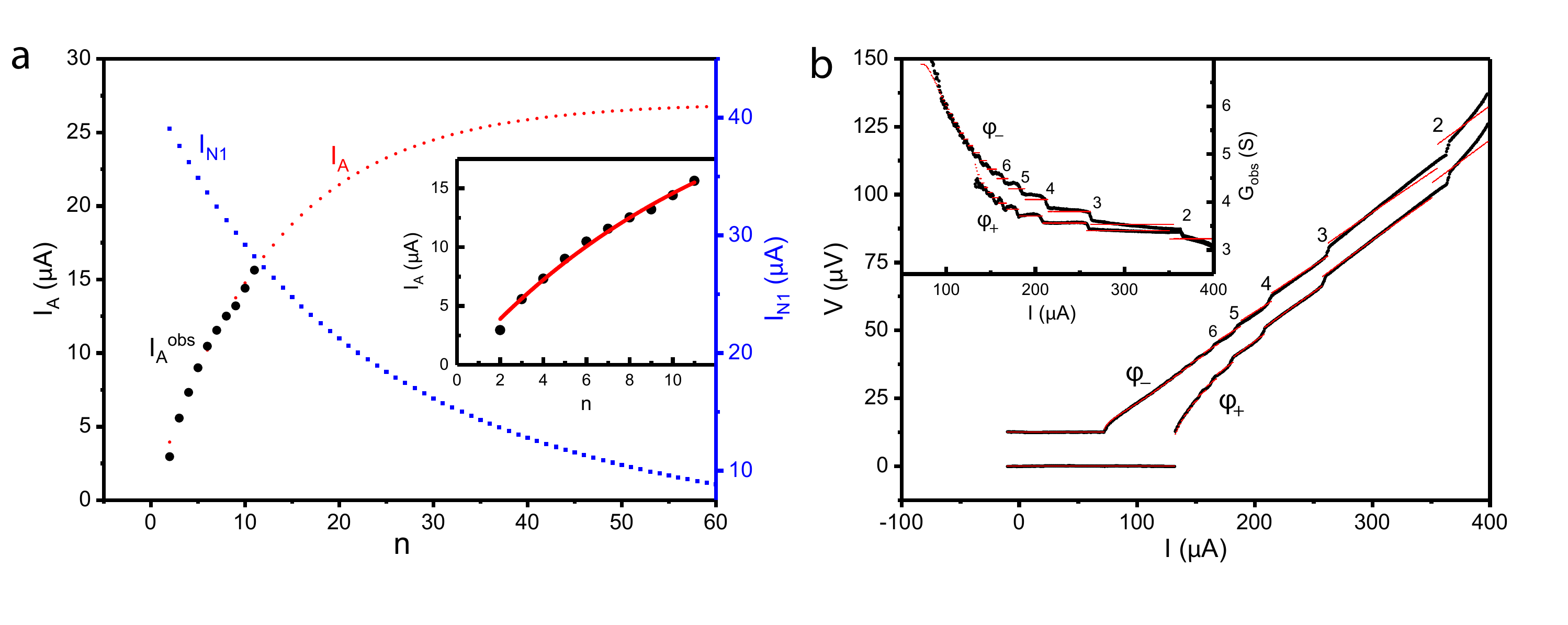}
\caption{\label{figFit} 
}
\end{figure*}

\vspace{2mm}\noindent
{\bf Figure S2}\\\noindent
Panel (a) displays $I$-$V$ curves with $B\in [0, 42] \;\mu$T in steps of $\Delta B = 2\;\mu$T (complement of Fig. \ref{figVmap}d). The integers $n$ at each step is indicated.  Panel (b) plots the $T$ dependence of $V_{\rm gap}$ inferred from the MAR sequence at each $T$. The $T$ dependence is consistent with the gap profile in Al. Panel (c) shows the plot of $V_n$ vs. $1/n$. The linear relationship $V_n = V_{\rm max}/n$ confirms the identification of the steps with MAR. Panel d: The variation of the $S$-$N$-$S$ critical current $I_{\rm c}$ vs. junction spacing $d$ at 135 mK (red symbols) and 340 mK (black). At 135 mK, $I_{\rm c}$ asymptotes to $\sim 30\;\mu$A  as $d$ decreases below 200 nm.

\newpage

\setcounter{figure}{0}
\renewcommand{\thefigure}{S\arabic{figure}}

\begin{figure}
\includegraphics[width=9 cm]{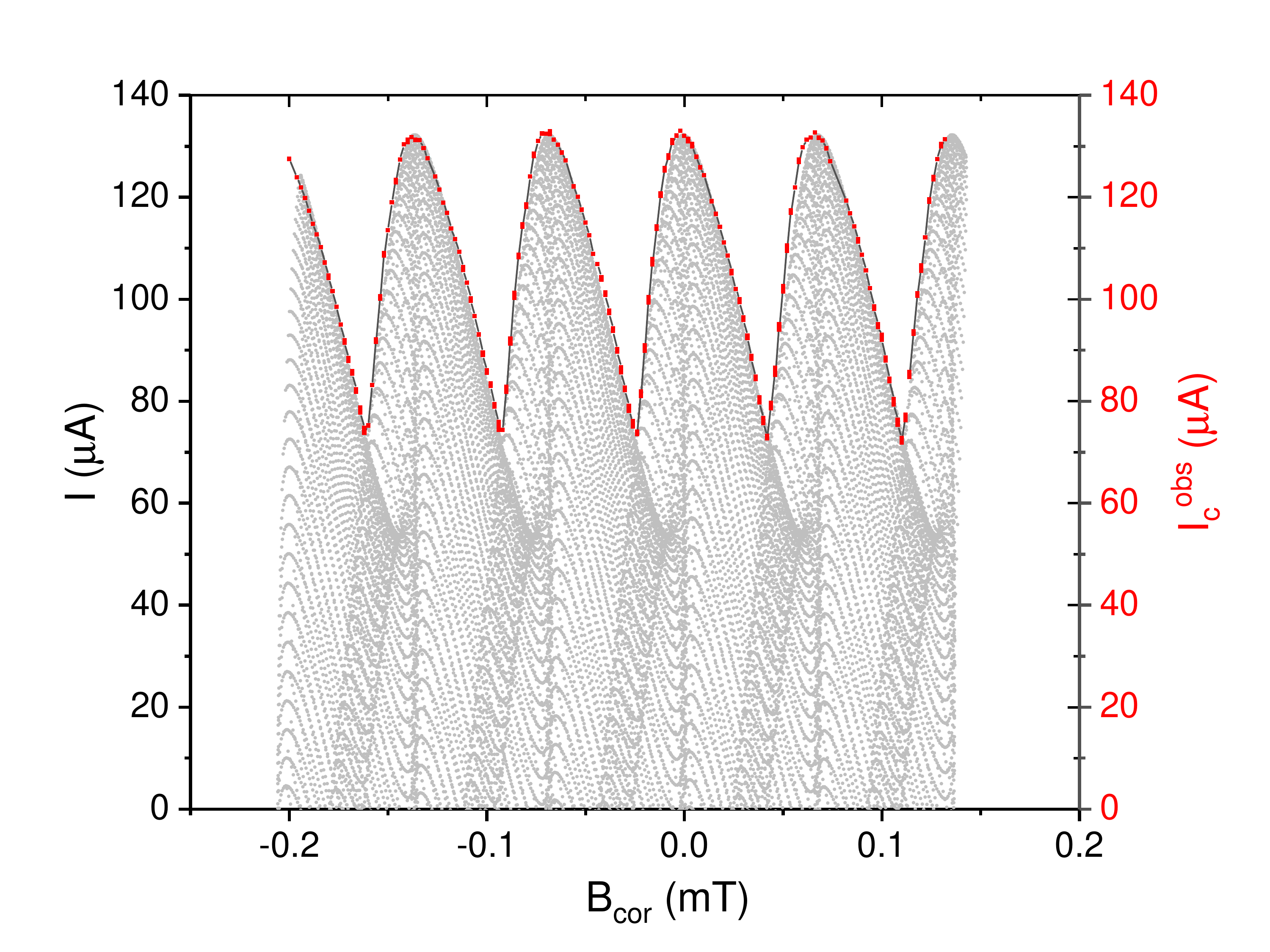}
\caption{\label{figCPR}
}
\end{figure} 

\begin{figure*}
\includegraphics[width=16 cm]{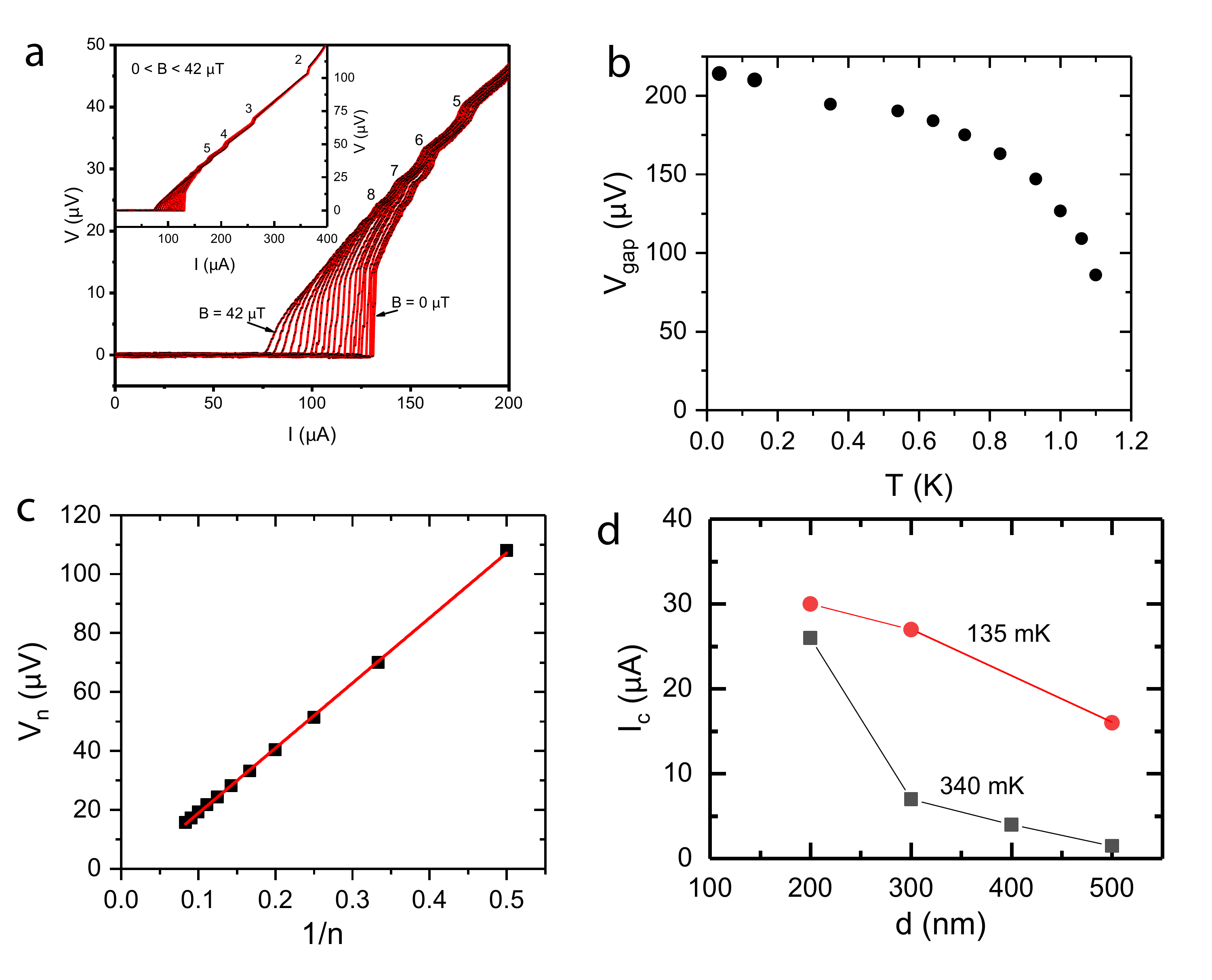}
\caption{\label{figAppendix}
}
\end{figure*}



\begin{thebibliography}{10}

\bibitem{Andreev} A. F. Andreev,
The Thermal Conductivity of the Intermediate State in Superconductors,
\emph{Zh. Eksp. Teor. Fiz.} {\bf 46}, 1823 (1964) [Sov. Phys. JETP {\bf 19}, 1228 (1964)].

\bibitem{Herrero}
Hubert B. Heersche, Pablo Jarillo-Herrero, Jeroen B. Oostinga, Lieven M. K. Vandersypen and Alberto F. Morpurgo,
Bipolar supercurrent in graphene,
\emph{Nature} {\bf 446}, 56-59 (2007).
doi:10.1038/nature0555

\bibitem{DuAndrei}
Xu Du, Ivan Skachko, and Eva Y. Andrei,
Josephson current and multiple Andreev reflections in graphene SNS junctions,
\prb {\bf 77}, 184507 (2008).
DOI: 10.1103/PhysRevB.77.184507

\bibitem{Pillet}
J-D. Pillet, C. H. L. Quay, P. Morfin, C. Bena, A. Levy Yeyati and P. Joyez,
Andreev bound states in supercurrent-carrying carbon nanotubes revealed,
\emph{Nature Physics} {\bf 6}, 965-969 (2010).
DOI: 10.1038/NPHYS1811

\bibitem{Esteve}
M. Zgirski, L. Bretheau, Q. Le Masne, H. Pothier, D. Esteve, and C. Urbina,
Evidence for Long-Lived Quasiparticles Trapped in Superconducting Point Contacts,
\prl {\bf 106}, 257003 (2011).
DOI: 10.1103/PhysRevLett.106.257003

\bibitem{Pothier}
L. Bretheau, C. O. Girit, C. Urbina, D. Esteve, and H. Pothier
Supercurrent Spectroscopy of Andreev States,
\emph{Phys. Rev. X} {\bf 3}, 041034 (2013)
DOI: 10.1103/PhysRevX.3.041034



\bibitem{Strambini}
E. Strambini, S. D’Ambrosio, F. Vischi, F. S. Bergeret, Yu. V. Nazarov and F. Giazotto,
The $\omega$-SQUIPT as a tool to phase-engineer Josephson topological materials,
\emph{Nature Nanotechnology} {\bf 11}, 1055–1059 (2016).
DOI: 10.1038/NNANO.2016.157



\bibitem{Bouchiat}
Anil Murani \etal,
Ballistic edge states in Bismuth nanowires revealed by SQUID interferometry,
\emph{Nat. Commun.} {\bf 8}, 15941 (2017).
doi: 10.1038/ncomms15941.

\bibitem{Yacoby}
Sean Hart, Hechen Ren, TimoWagner, Philipp Leubner, Mathias Mühlbauer, Christoph Brüne,
Hartmut Buhmann, Laurens W. Molenkamp and Amir Yacoby,
Induced superconductivity in the quantum spin Hall edge, 
\emph{Nat. Phys.} {\bf 10} 638-643 (2014).
DOI: 10.1038/NPHYS3036

\bibitem{Kim} Wudi Wang, Stephan Kim, Minhao Liu, F. A. Cevallos, R. J. Cava, N. P. Ong,
Evidence for an edge supercurrent in the Weyl superconductor MoTe$_2$,
\emph{Science} {\bf 368}, 534–537 (2020).
10.1126/science.aaw9270.


\bibitem{Furusaki} A. Furusaki, H. Takayanagi and M. Tsukada,
Theory of quantum conduction of supercurrent through a constriction,
\prl {\bf 67}, 132-135 (1991).


\bibitem{Beenakker} C. W. J. Beenakker and H. van Houten,
Josephson current through a superconducting quantum point contact shorter than the coherence length,
\prl {\bf 66}, 3056-3059 (1991).

\bibitem{Furusaki99} Akira Furusaki,
Josephson current carried by Andreev levels in superconducting quantum point contacts,
\emph{Superlattice and Microstructures} {\bf 25}, 809-818 (1999).


\bibitem{SanJose} Pablo San-Jose, Jorge Cayao, Elsa Prada and Ramon Aguado,
Multiple Andreev reflection and critical current in topological superconducting nanowire junctions,
\emph{New Jnl. Phys.} {\bf 15}, 075019 (2013).
doi:10.1088/1367-2630/15/7/075019

\bibitem{vonOppen} 
Yang Peng, Falko Pientka, Erez Berg, Yuval Oreg, and Felix von Oppen,
Signatures of topological Josephson junctions,
\emph{Phys. Rev.} B {\bf 94}, 085409 (2016).
DOI: 10.1103/PhysRevB.94.085409

\bibitem{Halperin}
Falko Pientka, Anna Keselman, Erez Berg, Amir Yacoby, Ady Stern, and Bertrand I. Halperin,
Topological Superconductivity in a Planar Josephson Junction,
\emph{Phys. Rev. X} {\bf 7}, 021032 (2017).
DOI: 10.1103/PhysRevX.7.021032


\bibitem{Taylor}
B. N. Taylor and E. Burnstein, 
Excess currents in electron tunneling between superconductors,
\prl~ {\bf 10}, 14 (1963).

\bibitem{Marcus}
S. M. Marcus, 
The magnetic field dependence of the 2$\Delta /n$ structure observed in Pb-PbO-Pb superconducting tunneling junctions,
\emph{Phys. Lett.} {\bf 19}, 623 (1966).

\bibitem{Klapwijk} T. M. Klapwijk, G. E. Blonder and M. Tinkham, 
Explanation of subharmonic energy gap structure in superconducting contacts,
\emph{Physica} {\bf 109-110B}, 1657-1664 (1982).



\bibitem{Averin}  D. Averin and A. Bardas,
ac Josephson Effect in a single quantum channel,
\prl {\bf 75}, 1831-1834 (1995).


\bibitem{Bratus} E. N. Bratus, V. S. Shumeiko, and G. Wendin,
Theory of subharmonic gap structure in superconducting mesoscopic tunnel contacts, 
\prl {\bf 74}, 2111-2113 (1995).

\bibitem{Cuevas} J. C. Cuevas, A. Martin-Rodero and A. Levy Yeyati,
Hamiltonian approach to the transport properties of superconducting quantum point contacts,
\prb {\bf 54}, 7366-7379 (1996).


\bibitem{Ouboter}
A. Th. A. M. De Waele and R. De Bruyn Ouboter,
Quantum intereference phenomena in point contacts between two superconductors,
\emph{Physica} {\bf 41}, 225 (1969).


\bibitem{Fulton}
T. A. Fulton, L. N. Dunkleberger, and R. C. Dynes
Quantum interference properties of double Josephson Junctions,
\prb {\bf 6}, 855 (1972).

\bibitem{Barone} A. Barone and G. Paterno, \emph{Applications and Physics of the Josephson Effect} (John Wiley and Sons, 1982), Ch. 12, p. 375.

\end{thebibliography}
\end{document}